\documentstyle[aps,preprint]{revtex}
\makeatletter
\def\ps@myheadings{\let\@mkboth\@gobbletwo
\def\@oddhead{\begin{minipage}[b]{\textwidth}
Hummer, Garde, Garc{\'{\i}}a, Pohorille {\&} Pratt: {\em Hydrophobic
interactions} \hfill Page \thepage\vspace*{0.3\baselineskip} \hrule
\end{minipage}}
\def\@oddfoot{}\def\@evenhead{\@oddhead}
\def\@evenfoot{}\def\sectionmark##1{}\def\subsectionmark##1{}}
\def\thebibliography#1{\section*{REFERENCES AND NOTES\@mkboth
 {REFERENCES}{REFERENCES}}\list
{\arabic{enumi}.}{\settowidth\labelwidth{[#1]}\leftmargin\labelwidth
 \advance\leftmargin\labelsep
 \usecounter{enumi}}
 \def\newblock{\hskip .11em plus .33em minus .07em}
 \sloppy\clubpenalty4000\widowpenalty4000
 \sfcode`\.=1000\relax}

\def\@cite#1#2{({#1\if@tempswa , #2\fi})} %

\makeatother

\begin{document}
\newcommand{\MU}{\Delta\!\mu^{ex}}
\newcommand{\AVE}{\left\langle n \right\rangle}
\newcommand{\MTWO}{\left\langle n^2 \right\rangle}
\newcommand{\VOL}{\Delta\!V}

\title{A Heuristic Molecular Model of Hydrophobic Interactions}

\author{Gerhard Hummer, Shekhar Garde,\footnote[2]{present address:
Department of Chemical Engineering, University of Delaware, Newark, DE
19716, USA} Angel E. Garc\'{\i}a, Andrew
Pohorille,\footnote[3]{Department of Pharmaceutical Chemistry, UCSF,
San Francisco, CA 94143; and NASA - Ames Research Center, MS-239-4,
Moffett Field, CA 94035-1000, USA}\\and Lawrence R.
Pratt\footnote[1]{To whom correspondence should be addressed.
LA-UR-95-3486}}

\address{Theoretical Division, Los Alamos National Laboratory, Los
Alamos, NM 87545, USA}

\date{\today}

\maketitle

\narrowtext
\thispagestyle{empty}

\begin{abstract} Hydrophobic interactions provide driving forces for
protein folding, membrane formation, and oil-water separation.
Motivated by information theory, the poorly understood nonpolar solute
interactions in water are investigated.  A simple heuristic model of
hydrophobic effects in terms of density fluctuations is developed.  This
model accounts quantitatively for the central hydrophobic phenomena of
cavity formation and association of inert gas solutes; it therefore
clarifies the underlying physics of hydrophobic effects and permits
important applications to conformational equilibria of nonpolar solutes
and hydrophobic residues in biopolymers.  \end{abstract}

\clearpage Protein folding and formation of lipid membranes and micelles
are driven by hydrophobic interactions.  Understanding the nature of
these interactions of nonpolar solutes in water and quantitatively
describing them has been the topic of numerous experimental and
theoretical studies \cite{Blokzijl:93}.  Despite these efforts, a
general, quantitative, molecular scale theory is not available.
Motivated by information theory, we study molecular scale hydrophobic
effects by exploring the thermodynamics of {\em hydration} and {\em
association} of model hard core solutes\cite{pratt:80} where otherwise
the only reliable calculational tools are elaborate computer
simulations.  As we will show, a simple theory based upon local density
fluctuations proves sufficient to describe qualitatively {\em and}
quantitatively the central hydrophobic phenomena.  The theory can be
applied easily to solutes of arbitrary shape and size, and permits study
of structural equilibria of nonpolar molecules and residues.

Among the previous theories of hydrophobic hydration, the scaled
particle model \cite{Pierotti:63,Reiss:59} initiated an important line
of development.  It utilizes asymptotic results in the small and large
solute limits where direct information on solvent correlations is not
required.  In the intermediate region and for phenomena such as solvent
separated hydrophobic interactions,\cite{franks,pc:77} molecular
correlations are important.  Stillinger \cite{Stillinger:73} and others
\cite{Pratt:92} partly filled these gaps, but much remains missing in
our knowledge of hydrophobic effects.  Our understanding of hydrophobic
interactions is particularly meager.  The available solubility or
``hydrophobicity'' models have not encompassed the molecular structural
features that give rise to stable free energy minima with water
separating hydrophobic groups --- the so-called solvent separated
hydrophobic interactions.  Thus, the hydrophobicity models have not been
justified by reproducing these interactions and to this extent must be
regarded as of unproven validity for conformational analysis of
biomolecular structure in solution.

Chemical potentials of nonpolar solutes in water are the quantities of
ultimate interest in the theory of hydrophobic hydration.  We shall
mainly be concerned with the excess chemical potential
$\Delta\!\mu^{ex}$ of a hard core solute in water.  Statistical
mechanics relates $\MU$ to the probability $p_0$ of finding an empty
volume, or {\em cavity}, of given size and shape in water at
equilibrium, $\MU = - k_B T \ln p_0$.  However, $p_0$ becomes
exceedingly small for larger cavities, and direct calculations of $p_0$
(e.g., by computer simulations using test particle insertion
\cite{Pratt:92}) become impractical.  Our goal here is to provide a
model to access this region of cavity sizes based upon accessible, even
if indirect, information.  The model we seek should not only extend the
range of sizes but also apply to non-spherical cavity shapes.  Complete
information on the binomial moments of the occupancy of a volume
identical to the cavity of interest would permit a construction of $p_0$
based upon the well known expression \cite{Reiss:59,Pratt:92,MM:41} that
exactly relates $p_0$ to the fluctuations in the number $n$ of solvent
molecules
in the cavity region:  \begin{eqnarray} p_0 & = & 1 + \sum_{j=1}^{\infty}(-1)^j
\left\langle \frac{n!}{j!(n-j)!}\right\rangle~.  \label{eq:psum}
\end{eqnarray} This formula is useful in guiding the analysis of
theories\cite{Stillinger:73} but becomes impractical for larger cavities
because of its sensitivity to missing information.  In those instances,
higher binomial moments cannot be neglected and that information is
unavailable.  We require an approach that is less sensitive to missing
information while exploiting information readily available for cavities
of {\em all} sizes and shapes.  These are the low order moments of the
number $n$ of solvent molecules in the cavity volume $\VOL$,
\begin{eqnarray} \AVE & = & \rho \VOL, \label{eq:m1}\\ \left\langle
n(n-1) \right\rangle & = & \rho^2 \int_{\VOL} d{\bf r} \int_{\VOL} d{\bf
r}' g(|{\bf r}-{\bf r}'|), \label{eq:m2} \end{eqnarray} where $g(r)$ is
the water oxygen pair correlation function and $\rho$ is the water
density.  Information theory \cite{jaynes} provides a paradigm for
exploiting such quantities.  We consider the set of probabilities $p_n$
of finding exactly $n$ solvent molecules in the cavity volume,
$\sum_{n=0}^{\infty}p_n=1$.  Observation of no solvent molecules in the
cavity region is then just one of the elementary events, and $p_0$ is
just one of the desired probabilities.  We formulate models of the
distribution $p_n$ that satisfy known moments, Eqs.~\ref{eq:m1} and
\ref{eq:m2}.  Minimizing an information measure permits the inference of
probabilities of maximum likelihood in a sampling experiment that
satisfies the moment information\cite{jaynes}.  However, the accurately
parabolic character of $\ln p_n$ for cavity volumes of small
molecular size, observed in computer simulations of water
\cite{Gaussian}, suggests an even simpler model:  We adopt the form
$p_n=\exp(\lambda_0+\lambda_1 n+\lambda_2 n^2)$ with ``Lagrange
multipliers'' $\lambda_0,\lambda_1,\lambda_2$ to be determined by the
three moment conditions $\sum_{n=0}^{\infty}p_n=1$,
$\sum_{n=0}^{\infty}n p_n=\AVE$, and $\sum_{n=0}^{\infty}n^2 p_n=\MTWO$.
{}From this we extract the $p_0$ that provides the desired thermodynamic
result.

Figure \ref{fig:sphere} shows the chemical potential for a spherical
cavity in water calculated from the model as well as from direct
cavity statistics. We find excellent agreement in the range
considered. This simple theory based entirely on the pair correlation
function of water is capable of reproducing the thermodynamics of
cavity formation in the region that is accessible to direct computer
simulations
\cite{Pratt:92,stillinger:93,head-gordon:94,head-gordon:95}.

So far, we considered the {\em hydrophobic hydration} of a single
cavity.  We now proceed to study {\em hydrophobic interactions}.  The
free energy associated with bringing together two hydrophobic solutes
corresponds to a distance dependent potential of mean force (PMF).  The
developed methodology provides us with the means of calculating this
PMF.  Given the first and second moment of the particle number
distribution, an approximate chemical potential ${\MU}(r)$ can be
determined for forming a cavity made of two spheres with given radius
$R$ and distance $r$ using the model.  The PMF is then defined as
$W(r)={\MU}(r) - \lim_{s\rightarrow\infty} {\MU}(s)$.

In the following, we study the association of two cavities of methane
size in water.  For the radius of the spheres, we have chosen a value of
$R=0.33$~nm.  This radius corresponds to the smallest distance where
methane-water pair correlations reach 1.0 in commonly employed models
\cite{Hummer:95:e}.  The cavity PMF is shown in Fig.~\ref{fig:PMF}.  As
a reference, the cavity potential produced by the molecular dynamics
simulation of Smith and Haymet \cite{Smith:93:c} is included.  We
obtained this by subtracting the bare potential from the PMF of methane
association.  Again, we find qualitative and quantitative agreement
between our simple model and elaborate computer simulations.  Our cavity
PMF shows a strongly favored region with overlapping cavities, separated
by a distinct barrier from a solvent separated, stable minimum at about
0.7~nm distance, in agreement with the computer simulations for methane
association.  In addition, we observe a very shallow third minimum at
1.1~nm distance.

\begin{figure}[h]
\caption{Torsional PMF of $n$-butane. Butane was modeled as four
spheres with radius $R=0.33$~nm, bond length 0.153~nm and tetrahedral
bond angle. The result of the fluctuation model (solid line) is
compared with the cavity PMF of Beglov and Roux (dotted line; from
Fig.~8 of Ref.~\protect\cite{Beglov:94}). The cis, gauche, and trans
rotational states are shown as inserts.}
\label{fig:butane}
\end{figure}
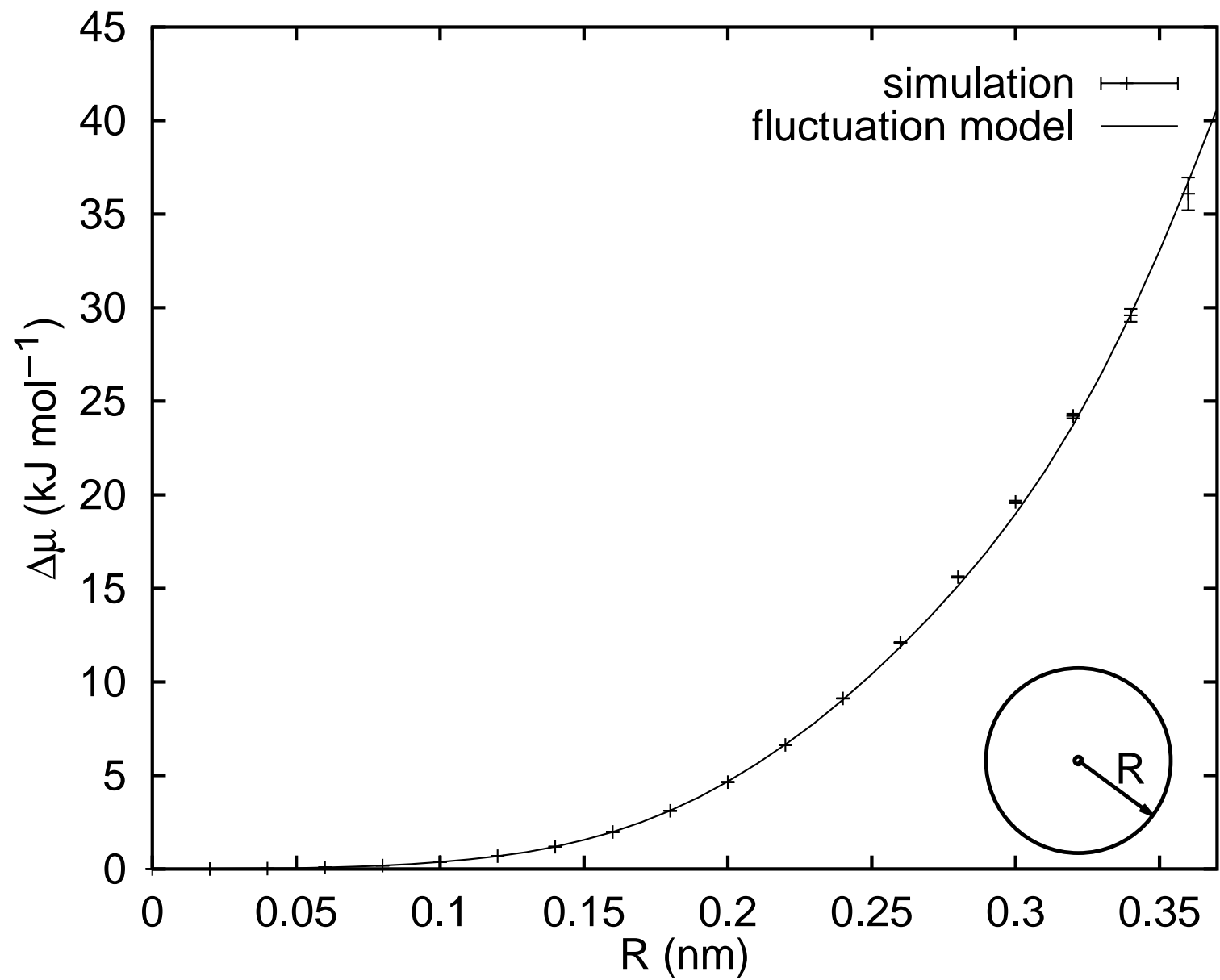
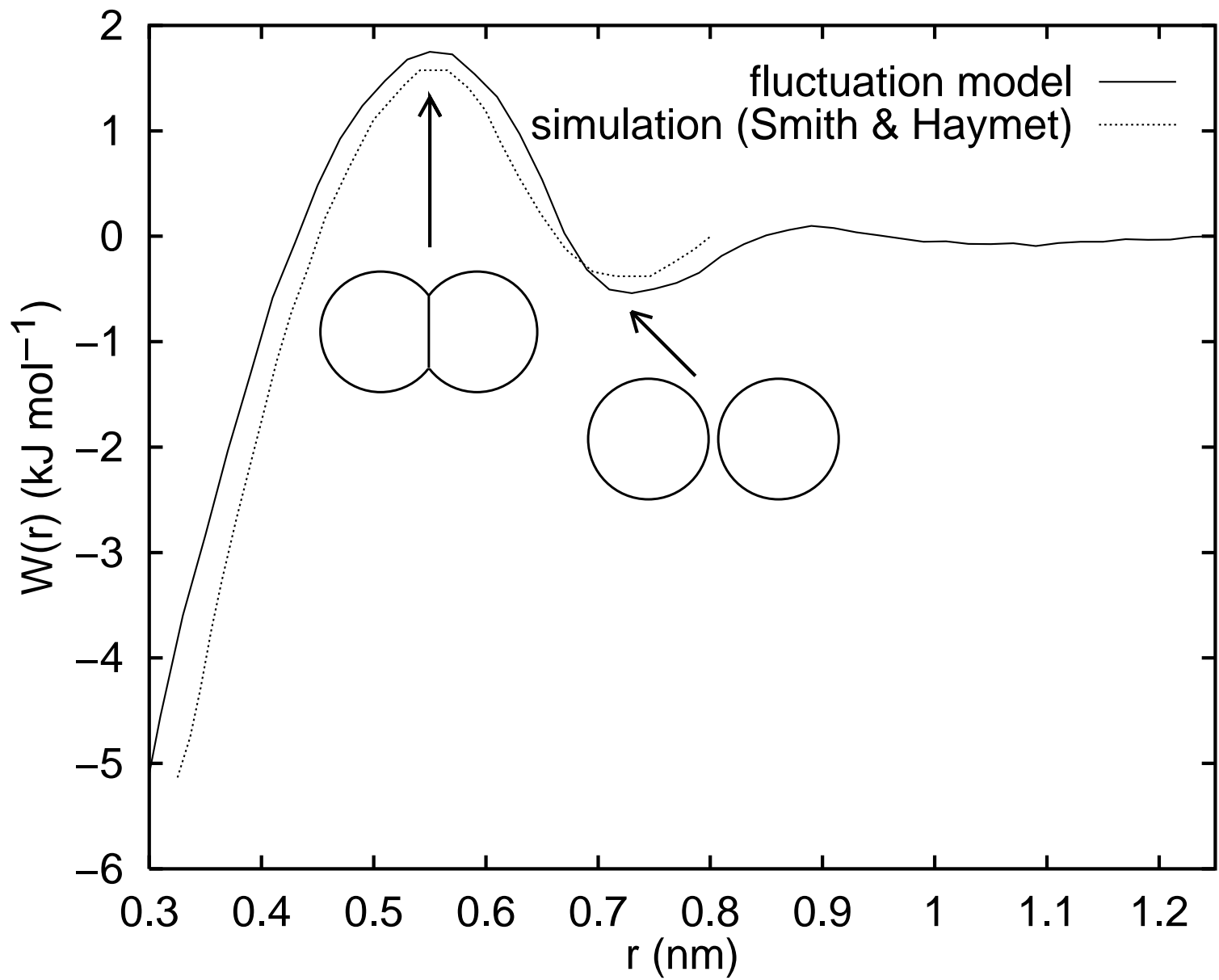
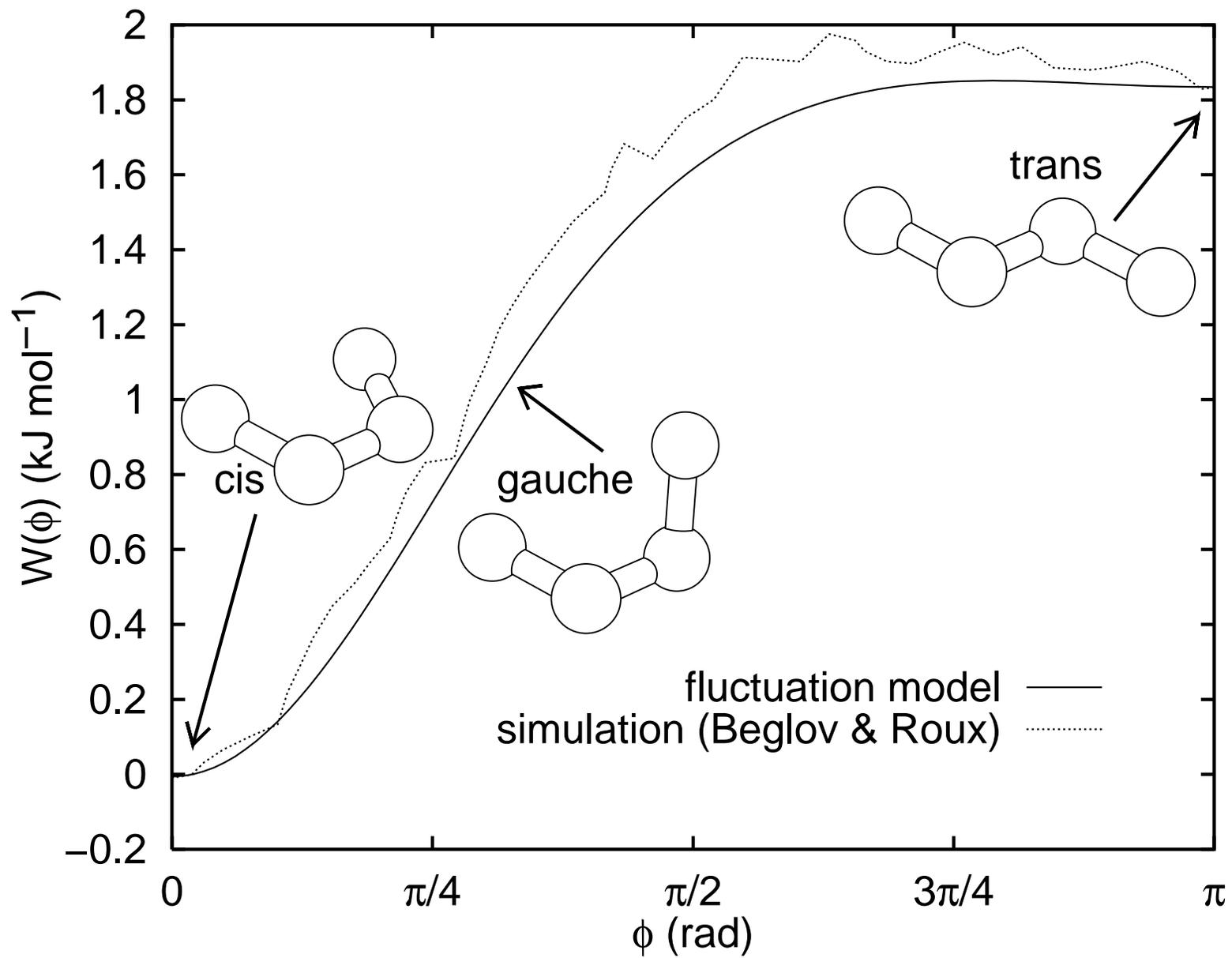

As a last example, we study the torsional equilibrium of $n$-butane.
Figure~\ref{fig:butane} shows the cavity PMF as a function of the
torsional angle $\phi$, which is compared to explicit
computer simulation results of Beglov and Roux \cite{Beglov:94}. In
complete agreement with the computer simulations, we find that the
more compact cis ($\phi=0$) and gauche structures ($\phi=\pi/3$) are
favored over the extended trans conformation ($\phi=\pi$) by about
1.8~kJ~mol$^{-1}$ and 0.7~kJ~mol$^{-1}$, respectively.

When the cavities of interest are much larger than the size of the
solvent molecules, other physical considerations intrude, as is well
known in the scaled particle developments\cite{Stillinger:73}.  We
expect this two moment model to become unsufficient as a sole
description in those cases.  Nevertheless, a broad range of problems of
biological interest, such as interactions of ligands with binding sites,
effects of point mutations on protein-solvent interactions, and
conformational equilibria of side chains, is within the range of
applicability of this model.

The present theory permits a sound, quantitative understanding of
hydrophobic hydration and association processes.  This view is based
upon (i)~the simplicity of the present theory involving only particle
number fluctuations, (ii)~the small amount of input information required
[$g(r)$] that is experimentally accessible, (iii)~the computational ease
of evaluating the theory permitting application to realistically large
and complex shaped solutes, (iv)~the unified and consistent treatment of
hydrophobic hydration and interaction phenomena, (v)~the clear
connection of information theory to basic statistical thermodynamics,
and (vi)~the qualitative and quantitative accuracy of the theoretical
predictions in a range that is otherwise inaccessible to direct
theoretical treatment.


\begin{thebibliography}{1}
\bibitem{Gaussian}
The relative incompressibility of water is reflected in a small
variance of particle-number fluctuations in macroscopic volumes, which
is related to the isothermal compressibility $\chi_T$ through
$\mbox{var}(n) = \langle n \rangle \rho k_B T \chi_T$. For an ideal
gas, $\rho k_B T \chi_T$ is 1; for water it is 0.06.
\end{thebibliography}

\begin{figure}[h] \caption{Hydrophobic hydration:  Comparison of the
chemical potential calculated from simulation (symbols) and the
fluctuation model (line) for spherical cavities with radius $R$, the
distance of closest approach of a water oxygen to the solute.  The
fluctuation integral Eq.~\protect\ref{eq:m2} required by the information
theory was reduced to one-dimensional integration
\protect\cite{Hill:58}.  The simulation results were gathered from
test-particle insertion, where 8000 configurations (separated by 50
passes each) of a Monte Carlo simulation of SPC water
\protect\cite{Berendsen:81} were used.  The same simulation was used to
calculate the water-oxygen pair correlation $g(r)$.}  \label{fig:sphere}
\end{figure}

\begin{figure}[h]
\caption{Hydrophobic interaction: PMF of cavity association. $r$ is the
distance of two spherical cavities with radius $R=0.33$~nm. The result
of the fluctuation model (solid line) is compared with the cavity PMF
of Smith and Haymet (dotted line; from Fig.~4 of
Ref.~\protect\cite{Smith:93:c}). The inserts illustrate the size of
the excluded volume. At the solvent-separated PMF minimum (0.7~nm), a
water molecule barely fits between the two hard-sphere solutes.}
\label{fig:PMF}
\end{figure}

\end{document}